# What is consciousness? Artificial intelligence, real intelligence, quantum mind, and qualia

STUART A. KAUFFMAN[1] and ANDREA ROLI[2,3]

[1]*Institute for Systems Biology, 401 Terry Ave N, Seattle, WA 98109, USA*

[2]*Department of Computer Science and Engineering, Alma Mater Studiorum Università di Bologna, Campus of Cesena, Via dell'Università 50, Cesena, Italy*

[3]*European Centre for Living Technology, Dorsoduro, 3911, 30123 Venezia VE, Italy*

Email: stukauffman@gmail.com, andrea.roli@unibo.it

## Abstract

We approach the question "What is Consciousness?" in a new way, not as Descartes' "systematic doubt", but as how organisms find their way in their world. Finding one's way involves finding possible uses of features of the world that might be beneficial or avoiding those that might be harmful. "Possible uses of X to accomplish Y" are "Affordances". The number of uses of X is indefinite (or unknown), the different uses are unordered, are not listable, and are not deducible from one another. All biological adaptations are either affordances seized by heritable variation and selection or, far faster, by the organism acting in its world finding uses of X to accomplish Y. Based on this, we reach rather astonishing conclusions: (1) Artificial general intelligence based on universal Turing machines (UTMs) is not possible, since UTMs cannot "find" novel affordances. (2) Brain-mind is not purely classical physics for no classical physics system can be an analogue computer whose dynamical behaviour can be isomorphic to "possible uses". (3) Brain mind must be partly quantum—supported by increasing evidence at 6.0 sigma to 7.3 sigma. (4) Based on Heisenberg's interpretation of the quantum state as "potentia" converted to "actuals" by measurement, where this interpretation is not a substance dualism, a natural hypothesis is that mind actualizes potentia. This is supported at 5.2 sigma. Then mind's actualizations of entangled brain-mind-world states are experienced as qualia and allow "seeing" or "perceiving" of uses of X to accomplish Y. We can and do jury-rig. Computers cannot. (5) Beyond familiar quantum computers, we discuss the potentialities of trans-Turing-systems.

# Introduction: The Issues

This short paper makes four major claims:

i. Artificial general intelligence is not possible.
ii. Brain-mind is not purely classical.
iii. Brain-mind must be partly quantum.
iv. Qualia are experienced and arise with our collapse of the wave function.

These are quite astonishing claims. Even the first claim is major. Artificial Intelligence (AI) has made tremendous achievements since its first steps in the fifties of the last century, with Turing introducing the main concepts and questions regarding computing machines (Turing, 1950), and the enthusiastic research plan of the Dartmouth research summer project (McCarthy *et al.*, 1955). Amazed by today's AI system capabilities, some await with fear replacement by intelligent robots. We hope to show that this is not possible for wonderful and fundamental reasons: the becoming of any world with an evolving biosphere of philosophic zombies, let alone conscious free will agents, is, remarkably, beyond any mathematics we know.

The pathway to this insight depends upon a prior distinction between the degrees of freedom in physics and in an evolving biosphere. In physics, the degrees of freedom include position and momentum, energy and time, the U(3)U(2)U(1) group structure of particle physics, the Schrödinger equation, General Relativity, and Dreams of a Final Theory (Weinberg, 1994; Kaku, 2021).

Oddly, in the evolving biosphere, "affordances" are the degrees of freedom. An affordance is "The possible use, by me, of X to accomplish Y." Gibson (1966) points out that a horizontal surface affords a place to sit. Affordances are both possibilities and constraints for the behaviour of organisms (Heras-Escribano, 2019; Campbell *et al.*, 2019; Jamone *et al.*, 2016). In evolution, an existing protein in a cell used to conduct electrons also affords a structure that can be used as a strut in the cytoskeleton or bind a ligand. Evolution proceeds by organisms "stumbling upon ever new affordances and 'seizing' them by heritable variation and natural selection". "Evolution tinkers together adaptive contraptions", as François Jacob said (Jacob, 1977).

Evolution has involved the evolution of behaviour. Thus, we wish to draw a fundamental distinction between "knowing" and "doing". Much of Western epistemology has centered on "knowing the world", but without "doing". This includes Descartes, Hume, Kant and Russell. But living organisms are non-equilibrium reproducing chemical reaction systems, hence to "survive" must obtain needed inputs. This requires ongoing interaction with the environment including other organisms. For each organism the relevant features of its environment, or *Umwelt*, (Uexkül, 2010), are the affordances, or opportunities and obstacles that "light up" in that world (Felin, *et al.*, 2021). Our new issue is, "how do we just "see" the old and novel affordances?

As a useful framework for the evolution of behaviour and, with it, mind, we borrow the U.S. military's phrase, "Observe, Orient, Decide, Act" or "OODA Loop". Of special interest is "Orient". To "orient" is to pick out relevant affordances in the current situation upon which to base action. How do organisms do this?

We humans do "just see" known and novel affordances. We easily do this when we tinker and jury-rig. Given a leak in the ceiling, we cobble together a cork wrapped in a wax-soaked rag stuffed into the hole in the ceiling and held in place with duct tape (Kauffman, 2019).

Jury-rigging uses subsets of the causal features of each object that articulate together to solve the problem at hand. Any physical object has alternative uses of diverse causal features.

An engine block can be used to drill holes to create cylinders and craft an engine, can be used as a chassis for a tractor, can be used as a paper weight, or its corners can be used to crack open coconuts (Kauffman & Roli, 2021).

**It is essential that there is no deductive relation between these uses. And there is, therefore, no deductive theory of jury-rigging (*ibid*.).**

How many uses of a screwdriver alone or with other things exist? Is the number exactly 16? No. Is the number infinite? How would we know? How define? No, the number of uses of a screwdriver alone or with other things is indefinite.

Consider some uses of a screwdriver alone or with other things. Screw in a screw. Open a can of paint. Scratch your back. Wedge a door closed. Scrape putty off the window. Tie to a stick and spear a fish. Rent the spear and take 5% of the catch …

What is the relation between these different uses? There are four mathematical scales, nominal, partial order, interval, ratio. The different uses of a screwdriver are merely a nominal scale. There is no ordering relation between the different uses of a screwdriver (Kauffman, 2019; Kauffman & Roli, 2021). This fact has profound and far-reaching consequences, as we show in the following.

## We Cannot Use Set Theory With Respect To Affordances

The Axiom of Extensionality states that: "Two sets are identical if and only if they have the same members." But we cannot prove that the indefinite and unlistable uses of a screwdriver are identical to the indefinite and unlistable uses of an engine block. No Axiom of Extensionality.

We cannot get numbers. One definition of the number "0" is "The set of all sets that have no element." This would be the set of all objects that have exactly 0 uses. Well, No. We cannot get the integers this way. We cannot get the number 1, or the number 17.

The alternative definition of numbers is via the Peano axioms. Define a null set, and a successor relation, N and N+1. But we cannot have a null set. And the uses of objects are unordered. There is no successor relation. We cannot get numbers from Peano. No integers, no rational numbers, no equations 2+3=5. No equations with variables 3+x=5. No irrational numbers. No real line. No equations at all. No imaginary numbers and no complex plane. No manifolds. No differential equations. No topology. No combinatorics and no first order predicate logic. No Quaternions, no Octonions. No "Well Ordering" so no "Axiom of Choice" so no taking limits (Kauffman & Roli, 2021).

A major implication affects computability: no non-embodied Universal Turing Machine (UTM), which operates algorithmically, hence deductively (Kripke, 2013), can find new affordances not already in its logical premises. In a computer program (we prefer the expression "computer program" over "algorithm" as it is more general and to emphasize the physical realization of a computational process), we represent the objects that are relevant for our purposes, along with their properties and their relations by means of a formal language. The program can then reason on this ontology and produce plans to solve a given problem. While doing this, both objects and relations can be combined by following constraints and rules in the knowledge base of the program. Nevertheless, a computer program cannot deduce new properties nor new relations. That is to say, the program cannot provide new explanations of the data it manipulates, besides the ones that can be deduced. The central reason is that, in general, there is no deductive relation between the uses of an object. From the use of an engine block as a paperweight a computer program cannot deduce its use as a way to crack open coconuts. It can of course find the latter use if it can be deduced, i.e. if there are: (A) a list of properties including the fact that the engine block has rigid and sharp corners, (B) a rule stating that one can break objects in the class of "breakable things" by hitting them against objects characterized by rigid and sharp corners, and (C) a fact stating that coconuts are breakable.

The universe of possibilities in a computer program is like a LEGO® bricks world: components with predefined properties and compositional relations that generate a huge space of possible combinations, even unbounded if more bricks can always be added. Now, let's suppose we add scotch tape, with which we can assemble bricks without being constrained by their compositional mechanism, and a cutter, which makes it possible to cut the bricks into smaller pieces of any shape. Here rules and properties are not predefined and we have a universe of indefinite possibilities: we are no longer trapped inside the realm of algorithms.

Besides deduction, other forms of logical reasoning exist, namely induction and abduction. The former proceeds from evidence to hypothesis: from the observation of black ravens, the hypothesis that "all ravens are black" is formulated. But the relevant "thing", raven, is already prestated. Induction is over already identified features of the world. Induction by itself does not identify new features of the world. Thus, it may be possible by induction to conclude that "all engine blocks can be used as paper weights", but from this induction it cannot be derived that the corners of the engine block can be used to crack open coconuts.

Abductive reasoning aims at providing an explanation of an observation by asserting a precondition that is likely to have this observation as a consequence. For example: if the corridor light bulb does not switch on, we can suppose it is broken. Abduction is differential diagnosis from a prestated set of conditions and possibilities. When implemented in computer programs, these kinds of reasoning nonetheless cannot add new symbols to represent new possibilities and new meanings. Abductive reasoning can only work with explanations already in its knowledge base. In other words, new symbols—along with their grounding in real objects—are outside of the ontology of the system.

The perhaps astonishing implication of this is that we humans and other organisms learn novel features of the world all the time, but cannot do so by deduction, induction or abduction with by using previous categories.

Our conclusion is supported by remarkable recent work (Devereaux, 2022) showing that "no modeler within the universe can have complete model of the universe". This work demonstrates that no finite list of true-false propositions and their truth values can exhaust the real world. New features of the world always exist and perhaps can be found and used. Open-ended behaving and doing in the world, therefore, cannot be limited to mere induction, deduction and abduction.

The further implications of this work, our own "The World Is Not a Theorem", (Kauffman, Roli, 2021), "The Third Transition in Science", (Kauffman, Roli, 2021), and recent results in the new field of Biocosmology indicating that the phase space of the biosphere has vastly expanded, (Cortês, *et al.*, 2022), and done so non-deductively, (Kauffman, Roli, 2021), are consistent with a view of co-evolving organisms mutually creating a diversifying web of mutual affordances, and thereby ever new ways of "getting to exist".

We can conclude from all of the above that non-embodied UTMs cannot find new affordances. Nor can such interacting Universal Turing Machines mutually create novel affordances. If finding and creating new affordances outside of the ontology of the UTM are necessary conditions for passing the Turing test, then non-embodied UTMs will never pass the Turing test. Moreover, besides the capability of reasoning and learning, an Artificial General Intelligence (AGI) should also be capable of using common sense knowledge, dealing with ambiguity and ill-defined situations, and creating new knowledge representations (Roli *et al.*, 2021). All these capabilities rely on the ability of finding affordances beyond the algorithmic predefined space, therefore AGI in non-embodied UTMs is ruled out.

Because affordances characterize actions in the physical world, a fundamental question arises as to whether and how robots, which are Embodied UTMs, can find and exploit novel affordances. Robots interact with the physical world through their sensors and actuators, and they can be capable of learning, therefore they can possibly discover new sensory-motor patterns useful for their goals. Nevertheless, two unresolved issues come into play: first, the symbol grounding problem (Harnad, 1990), i.e. how to attach new symbols to new sensory-motor patterns that reflect new features of the world. As stated by Harnad (Coradeschi, 2013), "sensory-motor transduction is not computation: it is physical dynamics". No algorithmic way of tackling this issue is therefore possible.

A second issue is the generalized version of the so-called frame problem (McCarthy & Hayes, 1969), i.e. the problem of specifying what is relevant for the robot's specific goal in this moment. This is the issue of "orient" in the OODA Loop. Consider a case of a robot using an engine block as a paper weight and the solution to achieving its goal is to use the engine block to crack open coconuts. To do so, the robot must acquire information on the relevant causal features of the engine block to crack open coconuts. The robot can move and sense its world via its sensors: What must occur such that the robot can discover the use of the engine block to crack open coconuts? **Achieving the final goal may require connecting several relevant coordinated causal features, none of which can be deduced from the others.**

For example, one way the robot might use the engine block to crack open coconuts is to rotate the engine block 40 degrees counterclockwise, tip the block to a 24 degree angle with respect to the floor, use its robotic arm to pick up an object, which actually is a coconut, and propel the object against one of the opposite corners of the engine block with some force, but stop propelling the coconut before destroying the coconut. It is clear that *indefinitely* many other ways to use the engine block to crack open coconuts also are possible, hence these are also affordances. **More, for any one such sequence of actions it is critical that there is no way for the robot to determine that it is actually improving over the successive and incremental steps of its search.** From the initial rotation of the engine block by 40 degrees, the robot cannot determine that this is, in fact, part of a possible multi-step solution to the non-deducible successive steps that ultimately succeed. The robot cannot accumulate successes until it happens upon the final success. **Even a first step is a search in an undefined space.** Taking this first step and each successive step to reaching the goal is blind luck with some time scale for each or perhaps many steps.

The example of the robot and engine block, real physical objects in the world, demonstrates that each "trial" by the robot in the real world requires a passage of finite interval of time. A passage of time is required because the robot must use the real physical object if it is to discover new novel but non-deducible features of the object. Thus, discovering a useful but non-deducible complex sequence of "actions" is a blind search in an indefinite space of possibilities.

Therefore, we conclude that even an embodied UTM can rarely find a concatenated set of novel affordances on some very long time scale compared to the time available to the robot to accomplish the task. Therefore, neither non-embodied nor embodied UTMs can attain AGI.

Computers cannot jury-rig in novel ways. The evolving biosphere can and does jury rig in ever-creative ways by jury-rigged Darwinian Pre-adaptations such as the evolution of the swim bladder from the lungs of lungfish (Kauffman, 2019). Cells do thermodynamic work to construct themselves. The evolution of the biosphere is a progressive jury-rigged construction not an entailed deduction (Kauffman, 2019; Kauffman & Roli, 2021). The evolution of hominid technology for the past 2.6 million years is also one of unending non-deductive jury rigging, ten stones 2.6 million years ago exploding to billions of goods including the space station today (Koppl *et al*., 2021).

Life and mind are not algorithms. Robots will not replace us. We can just see or perceive affordances. We can see and laugh about using an engine block as a paper weight and also to crack open coconuts. *Thus, we are not merely disembodied or embodied UTMs*.

But we do perceive affordances? How can we possibly perceive or "see" affordances?

## Perhaps We Are Classical Analogue Computers

Classical physics analogue computers can be embodied as robots. Analogue computers compute by being isomorphic to that that which is modelled. But we cannot be classical analogue computers. The reason is unexpected: affordances are "possible uses of X to do Y". But these Possibles are ontologically real: Mutations in evolving organisms are often indeterminate quantum events. Before the evolution of the swim bladder from the lungs of lungfish, was it possible that such a preadaptation would occur? Of course, the swim bladder really did come to exist, but some relevant mutation(s) may not have occurred, so the swim bladder might not have come to exist. To "exist or not exist" is surely ontological! Five hundred thousand years prior to the evolution of the swim bladder there was "no fact of the matter" concerning whether it might or might not emerge. C.S. Peirce pointed out that Actuals obey Aristotle's law of the excluded middle and also of non-contradition, (Kauffman, Humanity in a Creative Universe, 2016). "X is simultaneously true and not true", is a contradiction. All of classical physics obeys the law of the excluded middle and the law of non-contradiction. Possibles do not obey the law of non-contradiction. "X is simultaneously possibly true and possibly false" is not a contradiction (Kauffman, 2019).

In evolution, affordances are about ontologically real "possible uses of X to do Y". This is also true in our seeing affordances in our immediate world. It is really true that it is possible to use the corners of an engine block to crack open coconuts. In technological evolution it is really true 5000 years ago that the cross bow might or might not come to exist. Five thousand years ago there was "no fact of the matter" concerning whether or not the cross bow might come to exist.

Astonishingly, this implies that no classical system can be an analogue computer for affordances. Affordances do not obey the law of the excluded middle or the law of non-contradiction, but all classical systems do obey these laws. Thus, **no classical system can be isomorphic to, hence model, affordances.**

The claim that no classical system can be an analogue model isomorphic to affordances seems to be new and must survive severe critique.

## Brain-Mind Is Quantum

This hypothesis that brain-mind is partly quantum has been and is widely discussed (von Neumann, 1955; Wigner & Margenau, 1967; Penrose, 1989; Shimoney, 1997; Svetlichny, 2011). Alternative different proposals for actual molecular mechanisms for such quantum behaviour have been suggested, (Eccles, 1989; Fisher 2015; Fisher, 2017).

We wish to pursue a different set of data to support the claim that brain-mind is partly quantum. There are, at present, two tested experimental lines of evidence that brain-mind is partly quantum.

**First**, there is growing and powerful evidence gathered independently over seven decades by many independent investigators that mind may well be quantum, seen in aberrant behaviour of quantum random number generators, telepathy, and precognition. The data with respect to quantum random number generators are this: given a massive public event such as the death of Nelson Mandela, will quantum random number generators around the world behave non-randomly? The data are objective. These publicly available data are confirmed at 7.3 sigma (Radin & Kauffman, 2021). Do these data support the possibility of a quantum mind? Yes, if we ask, "Are these physically possible?" If mind is quantum, spatial nonlocality (Aspect *et al.*, 1982) with its entanglement allows telepathy between entangled minds and psychokinesis for entangled mind and matter. Temporal nonlocality (Filk, 2013), less well established, allows precognition (ibid). These phenomena are physically allowed if mind is quantum. It is now time to take decades of data seriously and disconfirm or extend them.

**Second**, a particular interpretation of quantum mechanics was offered by Werner Heisenberg. The quantum states are potentia hovering ghost-like between an idea and a reality (Heisenberg, 1958). We here adopt Heisenberg's view. *Reality consists in ontologically real Possibles, Res potentia, and ontologically real Actuals, Res extensa, linked by measurement*. This interpretation explains at least five mysteries of quantum mechanics, including nonlocality, which way information, null measurement, and "no facts of the matter between measurements" (Kauffman, 2016; Kastner *et al.*, 2018) so may rightly be considered seriously. It is of fundamental importance that ***Heisenberg's interpretation of quantum mechanics is not a substance dualism so does not inherit the mind body problem arising due to a substance dualism*** (Kauffman, 2016; Kastner *et al.*, 2018, Radin and Kauffman, 2021). *Thus, the hypothesis that brain mind is partly quantum **allows a new prediction**: it suggests a natural role for mind* (Radin & Kauffman, 2021). **Mind collapses the wave function**, as von Neumann, Wigner and Shimony, hoped (von Neumann, 1955; Wigner & Margenau, 1967; Shimony, 1997, Chalmers, 1996, Svetlichny, 2011).

Remarkably, this testable hypothesis stands quite well confirmed. Radin and others have shown at 5.2 sigma across 28 independent experiments that a human can try to alter the outcome of the two slit experiment and succeed. The effect is weak but significant. 5.2 sigma is one in 50,000,000 (Radin & Kauffman, 2021). These results, if more fully confirmed, have at least three stunning implications.

*First*, if strongly confirmed, the results alter the foundations of quantum mechanics (von Neumann, 1955; Wigner & Margenau, 1967; Shimony, 1997). Mind can play a role in the becoming of the universe. Since Newton, such role has been lost.

*Second*, for the first time since Newton, a Responsible Free Will is not ruled out. In the deterministic world of Newton, Free Will is impossible. Given quantum mechanics, the result of an actualization or measurement outcome is ontologically indeterminate, but fully random. I have Free Will but no Responsible Free Will. If I can try to alter the quantum outcome and succeed, responsible free will is not ruled out. This, if true, is transformative.

Our considerations here echo and parallel those of John A. Wheeler (Wheeler, 1988; Nesteruk, 2013). Based on his confirmed "delayed choice" experiment, Wheeler concluded that ours is a "participatory universe" in which we ask question of nature and do so by specific free-will chosen actions that alter the behaviour of the universe.

We are glad of this concordance. With a responsible free will we are indeed beyond Compatabilism (McKenna & Coates, 2019).

# The Third Potential Implication May Be The Most Important:
# We Try To And Do Collapse The Wave Function To A Single State.
# We Experience That State As A Qualia

The evidence for quantum aspects of mind and our capacity to play a role in "collapsing" or actualizing the wave function invite a new hypothesis for how we see affordances that we cannot see as classical systems including classical Universal Turing Machines. **Our Brain-Mind entangles with the world in a vast, entangled superposition. We** *try to and do collapse the wave function to a single state. We experience that state as a qualia*. Qualia! – Why not?

At least four further lines of evidence support the hypothesis just above that qualia are associated with collapse of the wave function.

*First*, as David Chalmers points out (Chalmers, 1996), qualia are never superpositions. He suggests from this that consciousness plays some role in the collapse of the wave function, (Chalmers, 1996; Chalmers & McQueen, 2022) Svetlichny makes much the same suggestion on slightly different grounds. (Svetlichny, 2011). We agree.

*Second*, finding novel affordances that just light up is not deductive. *Collapse of the wave function is also not deductive. Our experienced qualia are not deductions*. Neither need ideas that pop into mind when the Muse calls be deductions. Sudden insight gained upon grasping the point of a metaphor or the meaning of an ostensive definition that grounds a new symbol is also not a deduction. Insight in doing mathematics is not deductive (Byers, 2010). Creativity is not deductive, it is insight (Koestler, 1964). We cannot find new features of the world by deduction, induction or abduction. Insight is required. Insight is not deductive.

*Third*, our experienced world is experienced as an integrated "whole". This is the "unity of consciousness". This can be understood if brain-mind entangles in a rich superposition with the world. "Entangled quantum variables" are not independent of one another, but massively correlated. Actualization of any one variable immediately alters the amplitudes of the other variables. Successive actualizations may well be integrated "snapshots" of related worlds.

Fourth, our analysis of the incapacity of UTMs and any classical system to see novel affordances has a further implication. The evolution of the biosphere with *zombie organisms* can only find new affordances by accidentally stumbling upon them and seizing them by heritable variation and natural selection. It works but *is slow*.

We now stress again a central point. The task of multistep jury-rigging is not a search on a defined or deducible landscape. The first step of many steps gives no local clue that it is on a pathway to ultimate success. Yet humans easily carry out multistep jury-rigging. How can we possibly do this? We propose that the brain-mind entangles widely with the world, then we collapse this superposition and *directly experience relevant whole, linked and integrated sets of affordances that light up in our Umwelt,* (Felin, 2021). *Were we to experience no coordinated qualia, how could we determine multistep success? How would the world "light up" with coordinated relevant possibilities? This issue truly indicates that we really do experience qualia. This is, we claim, a major point.*

Thus, if some living organisms are in fact sentient with qualia, then such organisms can literally perceive, "*search and see*" (Gibson, 1966) old and new affordances. More, if they are responsible and free-willed and have endogenous goals, they can choose to act to use the affordance seen to achieve a goal. Watch a cat and mouse near a low chest of drawers. The chest affords the mouse a hiding place. The chest threatens the cat with mouse escape. We do this all the time. So did *T. rex*. As Walsh points out, the triad "agent, goal, affordance" arise together, (Walsh, 2015). This parallels an organism that "sees the world, evaluates it 'good or bad for me', and acts, (Peil, 2014). Peil suggests that this triad constitutes the first "sense", the emotion of hedonic value.

The resulting selective advantage of mind rapidly seeing and assessing relevant affordances via experienced qualia due to mind actualizing quantum potentia and free will choosing and acting is enormous. Mind can have and did evolve with diversifying life and behaviours and played a large role in the evolution of life that was far more rapid than were organisms as philosophic zombies. Niche construction is just one major area in evolutionary biology in which purposive behaviour plays a major role (Odling Smee *et al.*, 2003; Noble, 2006). Purposive, insightful behaviours as in Caledonian crows (Taylor *et al.*, 2010) and proto- ethical behaviours in primates (De Waal, 1996) can have and did evolve. The hominid lineage can have and did evolve, with evolving culture and technology (Koppl *et al.*, 2021). Experimental physicists really do purposefully line up and adjust their equipment with detailed planning and inventiveness. We really did decide to go to the moon and Mars, innovated and constructed the rockets, left our spacecraft on those planets and altered the orbital dynamics of the solar system. We really are responsible.

## Relation To Established Neurodynamics

The classical brain is dynamically critical (Beggs, 2008; a system in a dynamically critical regime is poised at the edge of chaos: Roli *et al.*, 2018). Genetic regulatory networks are critical (Kauffman, 1993; Daniels *et al.*, 2018; Villani *et al.*, 2018). Criticality is magical classically with small stable attractors, maximum entropy transfer, monotonic increase in basin entropy with the number of variables N, a power law distribution of small and large avalanches of changing variable activities that allows local and more global coordination, and graceful evolution under change of connections and logic (Bornholdt & Kauffman, 2019; Krawitz & Shmulevich, 2007; Aldana *et al.*, 2007). Life is co-evolving self-constructing Kantian Wholes dynamically on the Edge of Chaos (Kauffman, 2020; Roli & Kauffman, 2020). Co- evolving organisms may co-evolve to mutual criticality to maximize diversity of coordinated activities (Hidalgo *et al.*, 2014).

Years of superb work in neuroscience models suggests an astonishing diversity of brain dynamics perceptual behaviours with a variety of non-linear mathematical models (Grossberg, 2021). These models are entirely classical physics. If the claim that no classical system can constitute an analogue model for novel affordances is correct, as we claim, then a new pathway to investigate is the possibility of extending classical dynamical models to Hilbert space and seek homologous quantum behaviours.

Such homologous behaviour may be possible. For example: can Brain-mind be partly quantum and dynamically critical? Maybe with more specific hypotheses, e.g. Quantum scars (Turner *et al.*, 2018): the wave function remains in the vicinity of the classical attractor. Does the wave function of a quantum/classical critical brain remain in the vicinity of classical critical attractors that are usually taken to store alternative content addressable "memories"? In this quantum case, repeated actualizations could create highly similar qualia. In short, can such quantum systems inherit the magic of classical critical systems? Perhaps. Very recently a formalism to study quantum Boolean networks has been published (Franco *et al.*, 2021). More generally, can we seek a mapping from well-studied classical neuro-dynamics to quantum models with homologous behaviours? It should be possible to study quantum analogues of quantum criticality and chaos without and with decoherence (Kauffman *et al.*, 2012; Vattay *et al.*, 2015).

## Possible Soft Matter Systems To Examine And Trans-Turing Systems

At present enormous effort is focused on quantum computers that must maintain quantum coherence until decoherence or measurement achieves a solution, often the minimum of a complex classical potential, representing the solution, then computation stops (Das & Chakrabarti, 2008).

We suggest a new alternative. Cells do not stop. There is abundant evidence for quantum biology (Ritz *et al*., 2003; Kauffman et al., 2012; Pauls *et al.*, 2013; Brookes, 2017). Work in the past half-decade has explored a Poised Realm hovering reversibly between quantum and classical behaviour (Kauffman *et al*., 2012; Vattay *et al.*, 2015). Small molecules, peptides and proteins at room temperature can be quantum ordered, critical or chaotic. Quantum criticality lies at the metal/insulator transition. Such systems have delocalized wave functions, conduct electrons very well, and have power law slow decoherence that may underlie quantum effects in biology (Kauffman *et al*., 2012; Vattay *et al.*, 2015). The Schrödinger equation does not propagate unitarily in the presence of decoherence (Kauffman *et al*., 2012; Vattay *et al.*, 2015). The dynamical and physical behaviours of such soft matter systems will be new.

Intracellular and intercellular protein–protein complexes may constitute such a new class of physically embodied soft matter whose dynamical behaviours and can be studied with molecular dynamics and Lattice Boltzmann methods for quantum and classical behaviours (Succi, 2018). More such soft matter systems might constitute "Trans-Turing-Systems" with their own new internal dynamical behaviours and receiving and outputting quantum, classical and poised realm variables (Kauffman *et al*., 2012; Vattay *et al.*, 2015). Trans-Turing Systems will be a new class of non-deterministic dynamical systems of largely unknown behaviours. Such systems should be constructible now. Trans-Turing Systems (ibid) are beyond UTMs whose fundamental limitations as purely deductive syntactic systems we have explored here. New realms may open up conceptually and technologically. Living cells may be Trans-Turing Systems. We may not be too far off from creating soft matter systems including evolvable protocells that could constitute Trans-Turing Systems.

## Conclusion

The hypothesis that qualia are the experience of the actualized wave function, even if sensible, raises major issues: Are all actualizations of quantum superpositions associated with qualia in some form of panpsychism? Does the Strong Free Will Theorem bear on this issue (Conway & Kochen, 2006)? When a human is in a coma or dreamless sleep, are there qualia? What is unconscious mind from whence the Muse? How could we possibly test the hypothesis?

Moral: Artificial Intelligence currently is wonderful, but syntactic and algorithmic. We are not merely syntactic and algorithmic. Mind is almost certainly quantum, and it is a plausible hypothesis that we collapse the wave function, and thereby perceive coordinated affordances as qualia and seize them by identifying, preferring, choosing and acting to do so. We, with our minds, play an active role in evolution. The complexity of mind and coordinated behaviours can have evolved and diversified with and furthered the complexity of life. At last, since Descartes' lost his *Res Cogitans*, Mind can act in the world.

Free at last.

**Competing Interests**

The authors have no competing obligations.

**Acknowledgements**

# References

**Aldana M, Balleza E, Kauffman SA, Resendiz O. 2007.** Robustness and evolvability in genetic regulatory networks. *Journal of Theoretical Biology* **245**: 433–448.

**Aspect A, Dalibard J, Roger G. 1982.** Experimental test of Bell's inequalities using time-varying analyzers. *Physical Review Letters* **49**(25): 1804.

**Beggs JM. 2008.** The criticality hypothesis: how local cortical networks might optimize information processing. *Philosophical Transactions of the Royal Society* A **366**(1864): 329–343.

**Bornholdt S, Kauffman SA. 2019.** Ensembles, dynamics, and cell types: revisiting the statistical mechanics perspective on cellular regulation. *Journal of Theoretical Biology* **467**: 15–22.

**Brookes JC. 2017.** Quantum effects in biology: golden rule in enzymes, olfaction, photosynthesis and magnetodetection. *Proceedings of the Royal Society* A **473**(2201): 20160822.

**Byers W. 2010.** *How mathematicians think*. Princeton: Princeton University Press.

**Campbell C, Olteanu A, Kull K. 2019.** Learning and knowing as semiosis: extending the conceptual apparatus of semiotics. *Sign Systems Studies* **47**(3/4): 352–381.

**Chalmers DJ. 1996.** *The conscious mind: in search of a fundamental theory*. New York: Oxford University Press.

**Chalmers, DJ, McQueen, K, Gao S (ed.). 2022.** *Consciousness and Quantum Mechanics*. Oxford University Press (forthcoming).

**Conway J, Kochen S. 2006.** The free will theorem. *Foundations of Physics* **36**(10): 1441–1473.

**Coradeschi S. 2013.** Interview with Prof. Stevan Harnad. *KI-Künstliche Intelligenz* **27**(2): 169–172.

**Cortês M, Kauffman SA, Liddle AR, Smolin L. 2022.** Biocosmology: biology from a cosmological persective. https://www.biocosmology.earth; arXiv:2204.09379 [physics.hist-ph]

**Daniels BC, Kim H, Moore D, Zhou S, Smith HB, Karas B, Kauffman SA, Walker SI. 2018.** Criticality distinguishes the ensemble of biological regulatory networks. *Physical Review Letters* **121**(13): 138102.

**Das A, Chakrabarti BK. 2008.** Quantum annealing and analogue quantum computation. *Reviews of Modern Physics* **80**(3): 1061.

**Devereaux A, Koppl R, Kauffman SA, Roli A. 2022.** An Incompleteness Result Regarding Within-System Modeling https://papers.ssrn.com/solKKi3/papers.cfm?abstract_id=3968077
**De Waal FBM. 1996.** *Good natured. The origins of right and wrong in humans and other animals*. Cambridge, MA: Harvard University Press.
**Eccles J. 1989** *Evolution of the Brain: Creation of the Self*. Routledge.
**Felin T, Kauffman SA, Zenger T. 2021.** Resource origins and search, *Strategic Management Journal* https://doi.org/10.1002/smj.3350
**Filk T. 2013.** Temporal non-locality. *Foundations of Physics* **43**(4): 533–547.
**Fisher MPA. 2015.** Quantum cognition: the possibility of processing with nuclear spins in the brain. *Annals of Physics* **362**: 593–602.
**Fisher MPA. 2017.** Are we quantum computers, or merely clever robots? *Asia Pacific Physics Newsletter* **6**(1): 39–45.
**Franco M, Zapata O, Rosenblueth DA, Gershenson C. 2021.** Random networks with quantum Boolean functions. *Mathematics* **9**(8): 792.
**Gibson JJ. 1966.** *The senses considered as perceptual systems*. Boston: Houghton Mifflin.
**Grossberg S. 2021.** *Conscious mind, resonant brain: how each brain makes a mind*. Oxford: Oxford University Press.
**Harnad S. 1990.** The symbol grounding problem. *Physica D: Nonlinear Phenomena* **42**(1/3): 335–346.
**Heisenberg W. 1958.** *Physics and philosophy: the revolution in modern science*. New York: Harper Torchbooks.
**Heras-Escribano M. 2019.** *The philosophy of affordances*. Cham, Switzerland: Springer.
**Hidalgo J, Grilli J, Suweis S, Muñoz MA, Banavar JR, Maritan A. 2014.** Information-based fitness and the emergence of criticality in living systems. *PNAS* **111**: 10095–10100.
**Jacob F. 1977.** Evolution and tinkering. *Science* **196**(4295): 1161–1166.
**Jamone L, Ugur E, Cangelosi A, Fadiga L, Bernardino A, Piater J, Santos-Victor J. 2016.** Affordances in psychology, neuroscience, and robotics: a survey. *IEEE Transactions on Cognitive and Developmental Systems* **10**(1): 4–25.
**Kaku M. 2021.** *The god equation: the quest for a theory of everything*. New York: Doubleday.
**Kastner R, Kauffman SA, Epperson M. 2018.** Taking Heisenberg's potentia seriously. *The Journal of Quantum Foundations* **4**: 158–172.
**Kauffman SA. 1993.** The origins of order: self-organization and selection in evolution. New York: Oxford University Press.
**Kauffman SA. 2016.** *Humanity in a creative universe*. New York: Oxford University Press.
**Kauffman SA. 2019.** *A world beyond physics: the emergence and evolution of life*. New York: Oxford University Press.
**Kauffman SA. 2020.** Answering Schrödinger's "What Is Life?" *Entropy* **22**(8): 815.
**Kauffman SA, Roli A. 2021.** The world is not a theorem. *Entropy* **23**(11): 1467.
**Kauffman SA, Roli A. 2021.** The third transition in science: beyond Newton and quantum mechanics. A Statistical Mechanics of Emergence.
https://osf.io/m9kpz and arXiv http://arxiv.org/abs/2106.15271
**Kauffman SA, Vattay G, Niiranen G. 2012.** Uses of systems with degrees of freedom poised between fully quantum and fully classical states. *Patent Application Publication*. Publication date: Mar. 22, 2012.
**Koestler A. 1964.** *The act of creation*. New York: Macmillan.
**Koppl R, Devereaux A, Valverde S, Solé R, Kauffman SA, Herriot J. 2021.** Explaining technology. *SSRN papers*. https://papers.ssrn.com/sol3/papers.cfm?_abstract_id=3856338, May 30 2021.

**Krawitz P, Shmulevich I. 2007**. Basin entropy in Boolean network ensembles. *Physical Review Letters* **98**(15): 158701 (4 pp.). doi: 10.1103/PhysRevLett.98.158701

**Kripke SA. 2013.** The Church-Turing "thesis" as a special corollary of Gödel's completeness theorem. *In* Copeland BJ, Posy CJ, Shagrir O (eds), *Computability: Gödel, Turing, Church, and beyond* (chapter 4). Cambridge, MA: MIT Press, pp. 77–104.

**Kull K. 2001.** Special Issue Jakob von Uexküll: A paradigm for biology and semiotics." *Semiotica* **134** (1/4). Berlin: Mouton de Gruyter.

**McCarthy J, Hayes PJ. 1969.** Some philosophical problems from the standpoint of artificial intelligence. *In*: Meltzer B, Michie D (eds), *Machine Intelligence 4*. Edinburgh: Edinburgh University Press, pp. 463–502.

**McCarthy J, Minsky ML, Rochester N, Shannon CE. 1955.** A proposal for the Dartmouth Summer Research Project on Artificial Intelligence, August 31, 1955. *AI Magazine* **27**(4): 12. doi: https://doi.org/10.1609/aimag.v27i4.1904

**McKenna M, Coates DJ. 2019.** Compatibilism. *In* Zalta (ed.), *The Stanford Encyclopedia of Philosophy*. Stanford: The Metaphysics Research Lab, Stanford University.

**Nesteruk AV. 2013**. A "Participatory Universe" of J. A. Wheeler as an intentional correlate of embodied subjects and an example of purposiveness in physics. *Journal of Siberian Federal University. Humanities & Social Sciences* 6(3):415–437.

**Noble D. 2006.** *The music of life*. Oxford: Oxford University Press.

**Odling Smee J, Laland K, Feldman M. 2003.** *Niche construction: the neglected process in evolution*. Princeton: Princeton University Press.

**Pauls JA, Zhang Y, Berman GP, Kais S. 2013.** Quantum coherence and entanglement in the avian compass. *Physical Review E* **87**(6): 062704. doi: https://doi.org/10.1103/PhysRevE.87.062704

**Peil, KT. 2014.** Emotion: the self-regulatory sense. *Global Advances in Health and Medicine* **3**(2): 80-108

**Penrose R. 1989.** *The emperor's new mind*. Oxford: Oxford University Press.

**Radin D, Kauffman SA. 2021.** Is brain-mind quantum? A theory and supporting evidence. *arXiv preprint* arXiv:2101.01538.

**Ritz T, Damjanovic A, Schulten K. 2003.** The quantum physics of photosynthesis. *Quantum* **243**: 248.

**Roli A, Jaeger J, Kauffman SA. 2022.** How organisms come to know the world: fundamental limits on Artificial General Intelligence. *Frontiers in Ecology and Evolution* **9**:806283 (14 pp.). https://doi.org/10.3389/fevo.2021.806283

**Roli A, Kauffman SA. 2020.** Emergence of organisms. *Entropy* **22**(10): 1163 (12 pp.). https://doi.org/10.3390/e22101163

**Roli A, Villani M, Filisetti A, Serra R. 2018.** Dynamical criticality: overview and open questions. *Journal of Systems Science and Complexity* **31**(3): 647–663.

**Shimony A. 1997.** *On mentality, quantum mechanics, and the actualization of potentialities*. Cambridge, UK: Cambridge University Press.

**Succi S. 2018.** *The lattice Boltzmann equation – for complex states of flowing matter*. Oxford: Oxford University Press.

**Svetlichny, G. 2011.** Qualia are quantum leaps. *arXiv*:1104.2634 [physics.hist-ph] https://doi.org/10.48550/arXiv.1104.2634

**Taylor AH, Elliffe D, Hunt GR, Gray RD. 2010.** Complex cognition and behavioural innovation in New Caledonian crows. *Proceedings of the Royal Society* B **277**(1694): 2637–2643.

**Turing AM. 1950.** Computing machinery and intelligence. *Mind* **59**(236): 433–460.

**Turner CJ, Michailidis AA, Abanin DA, Serbyn M, Papić Z. 2018.** Quantum scarred eigenstates in a Rydberg atom chain: entanglement, breakdown of thermalization, and stability to perturbations. *Physical Review* B **98**(15): 155134.

**Uexküll, J von. 2010.** *A foray into the worlds of animals and humans: With a theory of meaning.* Minneapolis, Minnesota: University of Minnesota Press. Trad. by J.D. O'Neil. Originally published in German in 1934.

**Vattay G, Salahub D, Csabai I, Nassimi A, Kaufmann SA. 2015.** Quantum criticality at the origin of life. *Journal of Physics: Conference Series* **626**: 012023. IOP Publishing.

**Villani M, La Rocca L, Kauffman SA, Serra R. 2018.** Dynamical criticality in gene regulatory networks. *Complexity* **2018**: 5980636 (14 pp.). https://doi.org/10.1155/2018/5980636

**von Neumann R. 1955.** *Mathematical foundations of quantum mechanics*. Princeton: Princeton University Press.

**Walsh DM. 2015.** *Organisms, agency, and evolution*. Cambridge: Cambridge University Press. doi: 10.1017/CBO9781316402719

**Weinberg S. 1994.** *Dreams of a final theory*. New York: Vintage.

**Wheeler JA. 1988.** World as a system self-synthesized by quantum networking. *IBM Journal of Research and Development* 32(1):4–15.

**Wigner E, Margenau H. 1967.** Remarks on the mind body question. Symmetries and reflections. *American Journal of Physics* **35**(12): 1169–1170.